\documentclass[twocolumn,prb,superscriptaddress]{revtex4}
\usepackage{graphicx}
\usepackage{amsmath}

\usepackage{amsfonts}
\usepackage{amssymb}
\usepackage{bm}

\begin{document}


\title{Silver-Epoxy Microwave Filters and Thermalizers for Millikelvin Experiments} 
\author{Christian P. Scheller}
\author{Sarah Heizmann}
\author{Kristine Bedner}
\author{Dominic Giss}
\affiliation{University of Basel, Klingelbergstrasse 82, CH-4056 Basel, Switzerland}
\author{Matthias Meschke}
\affiliation{Low Temperature Lab, Aalto University, School of Science, 00076 Aalto, Finland}
\author{\;\hspace{5cm}\;Dominik M. Zumb\"uhl}
\email{dominik.zumbuhl@unibas.ch}\affiliation{University of Basel, Klingelbergstrasse 82, CH-4056 Basel, Switzerland}
\author{Jeramy D. Zimmerman}
\affiliation{Materials Department, University of California, Santa Barbara, CA93106, USA}
\author{Arthur C. Gossard}
\affiliation{Materials Department, University of California, Santa Barbara, CA93106, USA}



\date{\today}

\begin{abstract}
We present Silver-epoxy filters combining excellent microwave attenuation with efficient wire thermalization, suitable for low temperature quantum transport experiments. Upon minimizing parasitic capacitances, the attenuation reaches $\geq$\,100\,dB above $\mathrm{\approx150\,MHz}$ and - when capacitors are added - already above $\mathrm{\approx30\,MHz}$. We measure the device electron temperature with a GaAs quantum dot and demonstrate excellent filter performance. Upon improving the sample holder and adding a second filtering stage, we obtain electron temperatures as low as 7.5\,$\pm$\,0.2\,mK in metallic Coulomb blockade thermometers.
\end{abstract}


\maketitle 


Advancing to lower system temperatures is of fundamental interest, allowing to resolve smaller energy scales and hence opening the door for new physics. Examples include fractional quantum hall states\cite{willett_1987,pan99,radu08,dolev08}, nuclear spin phase transitions\cite{Simon08,Braunecker09prl,Braunecker09,Scheller14}, and many more. Dilution refrigerators with base temperatures as low as $\mathrm{5\,mK}$ are readily available, but typical electron temperatures $T_\mathrm{e}$ in semiconductor nanoscale devices are often considerably higher. This is due to insufficient thermalization, absence of adequate filtering of high frequency radiation\cite{Hergenrother95,Pekola10} as well as low frequency noise from the measurement setup including ground loops.

To attenuate microwave radiation, various types of cryogenic filters have been developed, such as metal powder filters\cite{martinis87,fukushima97,milliken07,lukashenko08}, micro fabricated filters\cite{vion95,coutois95,sueur06,longobardi13}, thermo-coax cables\cite{zorin95,glattli97}, copper tapeworm filters\cite{spietz06,bluhm08}, thin film filters\cite{jin97} and lossy transmission lines\cite{slichter09}. An overview of various microwave filters is given in Ref.\,\onlinecite{bladh03}. Despite decades of research, the state of the art is not satisfactory for cooling to 10\,mK or below. Besides filtering strategies, reduction of internal heat leaks and proper thermalization for sample, sample holder and wires are equally important. We note that the electron-phonon coupling has a very strong temperature dependence\cite{pobell_1992,pan99}, scaling as $\mathrm{T^{5}}$, making low temperature cooling very challenging. Wiedemann-Franz cooling\cite{pobell_1992} through the electrical sample wires scales much more favorably, as $T^2$, and eventually becomes the dominant cooling mechanism\cite{pan99}.

In this Letter, we present Ag-epoxy cryogenic filters uniting excellent microwave attenuation with efficient thermalization, suitable for low temperature electronic measurements. The filters are modular, of small size, robust against thermal cycling, and possess a predictable attenuation spectrum which can be understood as a skin-effect filter in a lossy transmission line model. Thermometry experiments using a GaAs quantum dot and metallic Coulomb blockade thermometers (CBTs) are employed to investigate the filter performance, demonstrating CBT electron temperatures\cite{Casparis2012} down to 7.5\,$\pm$\,0.2\,mK.

\begin{figure}[b]
\centering\vspace{-3mm}
\includegraphics[width=8.2cm]{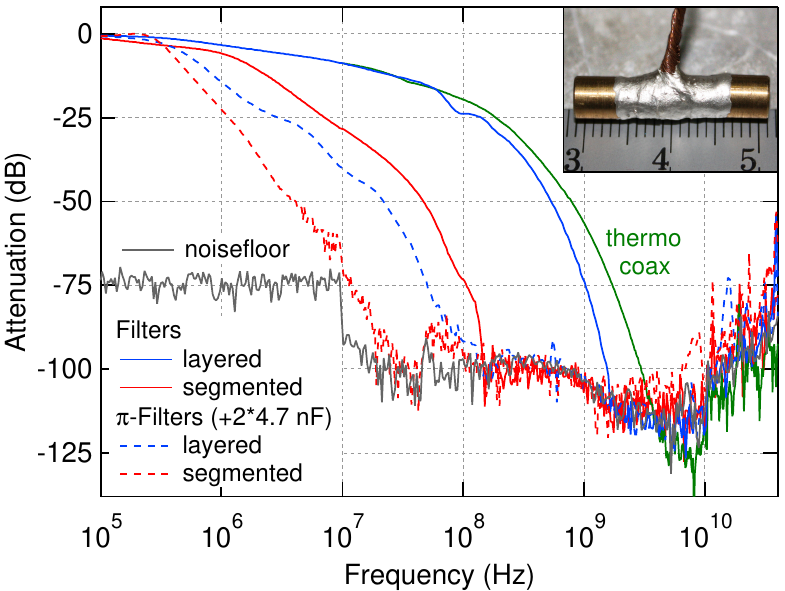}\vspace{-2mm}
\caption{Attenuation of thermo-coax and various filters at room temperature. Blue and red represent layered and segmented filters, respectively (see text), with DC resistance $\approx$\,5\,$\Omega$ ($<$\,50\,m$\mathrm{\Omega}$ at 4.2\,K) and capacitance $\approx$\,4\,nF. The dashed cures are from $\pi$-type filters with 4.7\,nF added at each end. The jump in the noise floor at 10\,MHz is due to use of a different analyzer at frequencies $\geq\,$10\,MHz. Cooling to 4.2\,K does not significantly alter the attenuation. Inset: photo of a filter (22\,mm long, 5\,mm diameter) with centimeter scale bar.}\vspace{-2mm}
\label{fig1}
\end{figure}

The filters are made by neatly winding 2.5\,m of insulated Cu-wire (diameter $D=\mathrm{0.1}$\,mm) with a magnet winding machine around a precast Ag-epoxy\cite{Ag_epoxy} rod to form a 5-layer coil. While winding, we continuously wet the rod and wires with Ag-epoxy to embed the wire in a conducting matrix, ensuring full electrical encapsulation and good thermal contact of the wire throughout the filter. Standard MCX connectors are soldered and, after gluing parts together with insulating epoxy\cite{Black_epoxy}, a high-conductivity Cu-braid is fastened to the filter also with Ag-epoxy, serving as a thermal anchor. A completed filter is shown in Fig.\,\ref{fig1} (inset). We mount the filters at the mixing chamber (MC) of a dilution refrigerator with $T_{\rm MC}\approx\,$5\,mK at base temperature. We emphasize that the wire is tightly embraced in the Ag-epoxy matrix, which is a good electrical conductor down to the lowest $T_{\rm MC}$. This is facilitating both cooling of the wire through the $d_\mathrm{Ins.}\sim$\,8$\,\mathrm{\mu m}$ thick polyurethane insulation as well as high frequency filtering -- thus presenting an advantage over earlier designs.

The attenuation profile is shown in Fig.\,\ref{fig1}, providing an attenuation $\geq$\,100\,dB above 1.5\,GHz (solid blue trace). Parasitic or stray capacitances can severely degrade the high-frequency attenuation. An optimized filter layout can reduce such capacitances: we split-up the coil into separate smaller segments arranged in series, resulting in a segmented filter, as opposed to the layered filter previously described. For layered filters, the first and the last layer of the coil are relatively close, separated only by 3 layers of Cu-wire or equivalently $\mathrm{0.3\,mm}$ distance. In contrast, the first and last coil are at a distance of $\mathrm{\approx5\,mm}$ for the segmented filters. In a naive plate capacitor model, this corresponds to a reduction of capacitive coupling by a factor of $\sim$\,10. Indeed, we obtain 100\,dB or more already above 150\,MHz for the segmented filters (red solid trace, Fig.\,\ref{fig1}), at 10 times lower frequencies compared to the layered filter.

The filter attenuation can be modeled as a lossy, distributed transmission line with skin-effect: the resistance per length $R_\mathrm{IC}$ and $R_\mathrm{OC}$ of inner and outer conductor, respectively, depend on frequency. While DC-currents are supported by the whole cross section $A=\pi D^{\,2}/4$ of the Cu-wire with diameter $D$, the skin effect forces AC-currents at frequency $\nu$ to an annulus of width $\delta=1/\sqrt{\sigma_\mathrm{Cu}\mu_\mathrm{Cu}\nu\pi}$ -- the skin depth. Here, $\sigma_\mathrm{Cu}$ denotes the conductivity and $\mu_\mathrm{Cu}$ the magnetic permeability of the Cu-wire. The effective cross section therefore reduces to \mbox{$A=D\pi\delta$} and consequently $R_\mathrm{IC}$ increases as a function of $\nu$. Equivalent arguments hold for the outer conductor. Note that due to its smaller conductivity $\sigma_\mathrm{epoxy}\ll\sigma_\mathrm{Cu}$, the Ag-epoxy outer conductor is dominating the total resistance $R_\mathrm{tot}\left(\nu\right)=R_\mathrm{IC}\left(\nu\right)+R_\mathrm{OC}\left(\nu\right)$ at high $\nu\gtrsim 2\,$MHz.

The resulting attenuation is $\propto -\nu^{3/4}$ for low frequencies $2\pi\nu L\ll R_\mathrm{tot}\left(\nu\right)$ and $\propto -\nu^{1/2}$ for high frequencies in the GHz regime with a smooth transition in between. Here, $L=\mu_0/2\pi\ln\left((D+2d_\mathrm{Ins.})/D\right)$ denotes the inductance per length. For a single layer filter, where additional parasitic capacitances not included in the theory are negligible, the model is in very good agreement with the measured attenuation without using any fit parameters, see supplementary material\cite{supplementary} for details. This demonstrates the validity of the skin-effect lossy transmission line model, and can explain the general functional shape of the measured filter attenuation curves. When comparing the model with the segmented and layered filters described above, the agreement between theory and measured attenuation is still good once a fit parameter is allowed, e.g. the capacitance per length.

A $\pi$-type filter can be created by adding capacitors\cite{pacificaerospace} to both filter ends, further increasing the attenuation\cite{lukashenko08}. Both layered and segmented $\pi$-filters are shown in Fig.\,\ref{fig1} as dashed curves, delivering an attenuation $\geq$\,100\,dB attenuation already above 30\,MHz for segmented $\pi$-type filter. However, in order not to reduce the bandwidth for future experiments too much and because of increased current preamplifier noise with large capacitances, the segmented filters without additional capacitors were chosen for the electron temperature measurements. Further, we note that replacing the central rod with a mixture of insulating epoxy\cite{Black_epoxy} and Fe powder\cite{lukashenko08} or replacing the copper wire with resistive wire did not improve the filter performance. Once properly completed, repeated thermal cycling has not yet caused failure of any of the over hundred filters in use in our laboratory.

We now turn to the filter performance measured by the device electron temperature $T_{\rm e}$ with a GaAs quantum dot thermometer\cite{Kastner92,Houten92,kouwenhoven96,Karakurt01,Potok07,Rossi12,Mavalankar,Torresani13,Maradan14}, comparing $T_{\rm e}$ with and without filters. For all measurements below, the sample wires are connected to the room temperature measurement setup through $\sim$1.5\,m long thermo-coax cables\cite{thermocoax}, which are very effective attenuators above a few GHz, see green curve in Fig.\,\ref{fig1}, up to very high frequencies $\gtrsim$\,1\,THz\cite{zorin95}. The Ag-epoxy filters provide additional thermalization and serve to significantly bring down the low-pass cut-off frequency of the combined thermo-coax and Ag-epoxy filter system.

\begin{figure}[t]
\centering
\includegraphics[width=8.2cm]{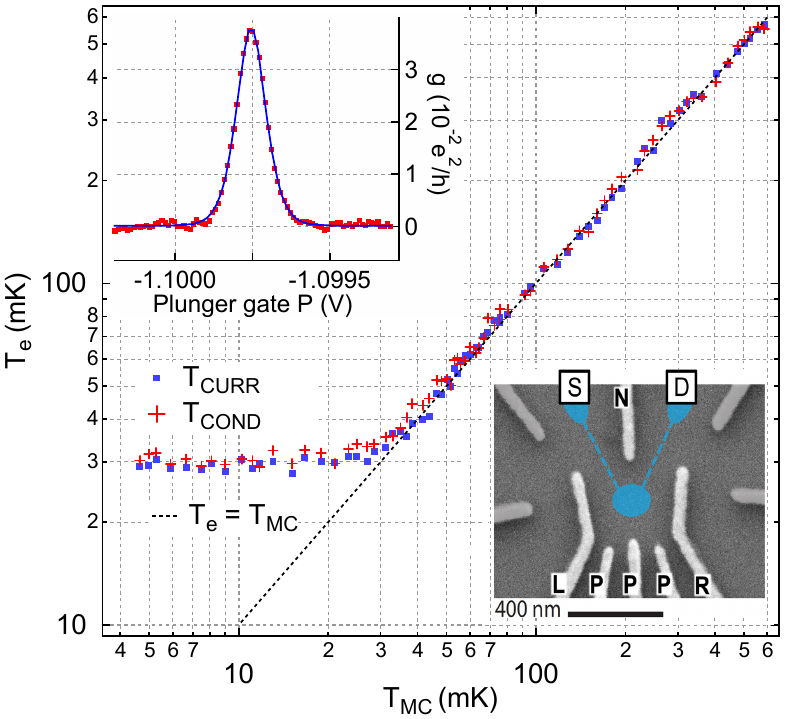}
\caption{Electron temperature $T_{\rm e}$ extracted from GaAs dot current and conductance measurements as indicated versus refrigerator temperature $\mathrm{T_{MC}}$, with dashed line showing ideal thermalization $\mathrm{T_\mathrm{e}=T_\mathrm{MC}}$. The upper inset shows a typical zero-bias peak at base temperature with $\cosh^{-2}$ fit\cite{Houten92}. Lower inset: micrograph of a similar device. Gates N,\,L,\,R and P (light grey) form the dot. Dark gates are grounded.} \vspace{-4mm}
\label{fig2}
\end{figure}

We use a surface gated GaAs/AlGaAs quantum dot (see Fig.\,\ref{fig2} lower inset) in the temperature broadened Coulomb blockade regime as an electron thermometer, probing the Fermi-Dirac distribution of the reservoirs. Gates N,\,L,\,R and P isolate the QD from the surrounding 2D gas, while source (S) and drain (D) ohmic contacts serve for current injection and detection. Details on sample fabrication are given in Ref.\,\onlinecite{zumbuhl06}. All measurements presented here were performed in the single electron regime. The dot differential conductance $g$ at zero DC-bias peaks whenever one of its energy levels is aligned with the chemical potential of source and drain, see upper inset, Fig.\,\ref{fig2}. The shape and width $\Delta V_\mathrm{G}$ in gate voltage of this peak reflects the reservoir temperatures and provides a primary electron thermometer\cite{Kastner92,Houten92,kouwenhoven96} in the limit of small tunnel rates to source and drain (and large dot level spacing) compared to temperature. The conversion between measured peak width $\Delta V_\mathrm{G}$ and energy $\Delta E$ is done by means of the lever arm $\alpha$, $\Delta E=\alpha\cdot\Delta V_\mathrm{G}$. To determine $\alpha$, a DC-bias $eV_\mathrm{SD}\gg k_B T_{\rm e}$ is applied (Boltzmann constant $k_B$ and electron charge $e>0$), splitting the zero-bias conductance peak into two peaks with gate-voltage separation $V_\mathrm{sep}$ and giving $\alpha=eV_\mathrm{SD}/V_\mathrm{sep}$. All conductance data is recorded with a standard lock-in technique using an AC-excitation experimentally chosen sufficiently small to avoid heating ($\mathrm{2\,\mu V}$ at base temperature).

Instead of differential conductance $g$ at $V_{\rm SD}=0$ we can also measure the DC dot current $I_{\rm DC}$ at large bias $eV_{\rm SD}\gg k_B T$, which shows transitions between low and high current when a dot level is crossing the chemical potential of source and drain. These current transitions reflect the Fermi-Dirac distribution of the corresponding reservoir, and allow extraction of source and drain temperature separately (see Ref.\,\onlinecite{Maradan14} for details). Here, $\alpha$ is simply given by the ratio of the applied bias $V_\mathrm{SD}$, divided by the separation of the inflection points for the two distributions. We stress that $I_{\rm DC}$ is determined by the tunneling rates and does not depend on the applied DC voltage as long as excited states (and cotunneling) do not contribute, as we assume here. Also, reducing the DC bias did not affect $T_\mathrm{e}$. Further, there is no systematic deviation between source and drain temperatures, and we conclude that energetic electrons at $\mathrm{\geq100\,\mu eV}$ become efficiently thermalized in the reservoirs.

Fig.\,\ref{fig2} displays the extracted electron temperatures from the same warmup using both $g$ and $I_{\rm DC}$. We measure the zero-bias $g$-peak with a finite AC excitation $V_{\rm AC}$ and then the $I_{\rm DC}$ steps at finite $V_\mathrm{SD}$, and repeat this cycle. We perform Fermi-Dirac distribution fits on the $I_{\rm DC}$ steps, forcing the two reservoir temperatures to be identical in order to obtain more reliable fitting results, and obtain the reservoir electron temperature $T_{\rm CURR}$ as well as the lever arm $\alpha$. Finally, we get the reservoir electron temperature $T_{\rm COND}$ from $\cosh^{-2}$ fits\cite{Kastner92,Houten92,kouwenhoven96} to the $g$-peak using the same lever arm. We note that for the $g$ measurements, already a small $eV_{\rm DC}\sim k_B T$ results in a broadened peak and an overestimation of $T_\mathrm{COND}$. Therefore, a feedback mechanism was implemented that compensates any $V_{\rm SD}$ drift by minimizing $I_{\rm DC}$ in the zero-bias measurement, therefore ensuring excellent alignment of source and drain during the $\sim$40 hour measurement shown in Fig.\,\ref{fig2}.

\begin{figure}[t]
\begin{tabular}{c}
\includegraphics[width=8.2cm]{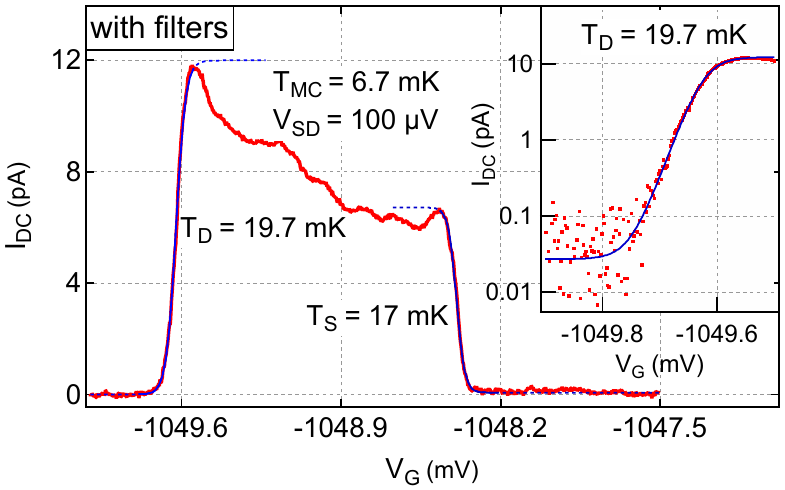} \\
\includegraphics[width=8.2cm]{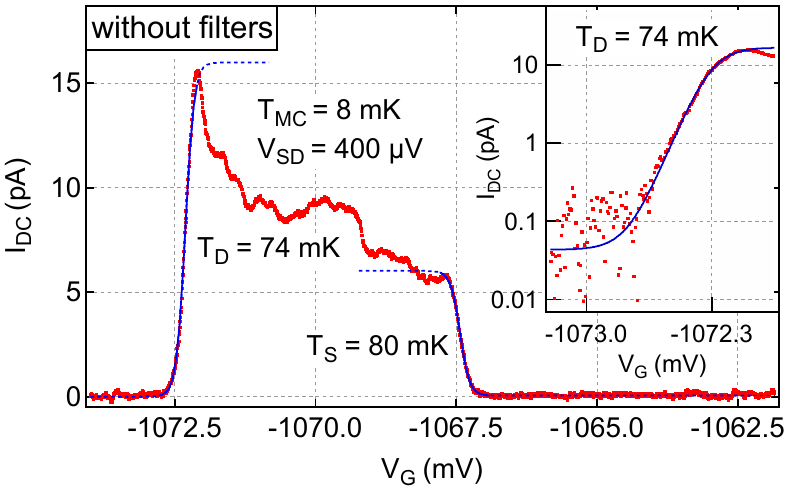} \\
\end{tabular}
\caption{DC current $I_{\rm DC}$ through the GaAs dot as a function of plunger gate voltage $V_{\rm G}$ taken at base temperature, with individual Fermi-Dirac fits for source and drain leads, as indicated, comparing data with filters (upper panel) and without filters (lower panel). The insets shows the same data and fit in log-scale for the drain side. The linear current increase seen here clearly indicates the temperature broadened regime.}\vspace{-4mm}
\label{fig3}
\end{figure}

Both electron temperatures $T_{\rm CURR}$ and $T_{\rm COND}$ agree well with each other over the whole temperature range and, above 40\,mK, also with $T_\mathrm{MC}$, measured with a $\mathrm{RuO_{2}}$ thermometer, thus demonstrating excellent thermometer operation. The $\mathrm{RuO_{2}}$ thermometer was precalibrated with a fixed point device and agrees very well with a CMN paramagnetic inductance thermometer\cite{pobell_1992}. Below 40\,mK, $T_\mathrm{e}$ starts deviating from $T_\mathrm{MC}$ and saturates at $\sim$30\,mK. Upon reducing electronic noise and simplifying the measurement setup (measuring only $I_{\rm DC}$), the electron temperature $T_\mathrm{e}$ is reduced down to $T_{\rm CURR}=\mathrm{18\pm3\,mK}$, obtained from an average over several traces. An example of such a trace including Fermi-Dirac fits of the current steps and individually extracted source/drain temperatures is presented in the upper panel of Fig.\,\ref{fig3}.

When the filters are removed (replaced with plain adapters) without changing any other part of the experiment, the electron temperature increases to $T_{\rm CURR}\approx$\,75\,mK, see lower panel of Fig.\,\ref{fig3}, clearly demonstrating the efficiency of the filters. Note that due to the elevated electron temperature, more DC bias is applied to better separate the source and drain current steps. However, this does not increase the electron temperature, as discussed already earlier (also here, smaller bias was tested).

There is still room for further improving the electron temperature, considering that the refrigerator cools to $T_{\rm MC}=$\,5\,mK. An additional miniaturized Ag-epoxy filtering/thermalization stage is placed directly inside the Faraday cup, i.e. inside the shielded sample can. Furthermore, a heat sunk sample holder made from conductive Ag-epoxy is used, and the ceramic chip carrier (known to suffer from heat release) is replaced with a standard plastic dip socket equipped with a 1\,mm thick gold plated and heat sunk copper backplane. We proceed to measure electron temperatures with this improved two stage filtering/thermalization setup. For simplicity, metallic CBTs are used that -- in contrast to GaAs dots -- do not require gate tuning to deliver the device electron temperature. We note that CBTs have cooled to very similar electron temperatures as the dot before improving the setup.

\begin{figure}[t]
\centering
\includegraphics[width=8.2cm]{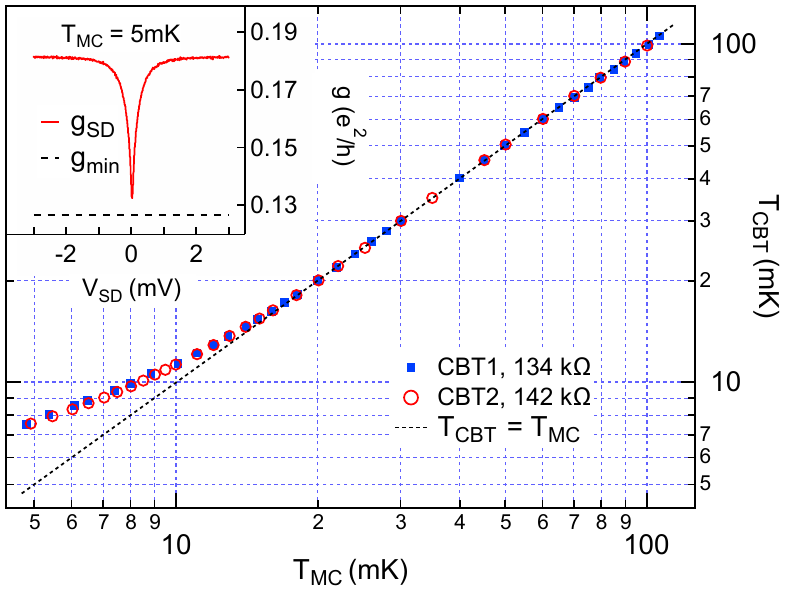}
\caption{CBT electron temperature $T_{\rm CBT}$ versus mixing chamber temperature $T_{\rm MC}$ for two CBTs on the same wafer, measured in a perpendicular magnetic field of 250\,mT to drive the Al normal. Ideal thermalization $T_{\rm CBT}=T_{\rm MC}$ is indicated as a dashed line. The inset shows $g$ as a function of $V_{\rm SD}$ at base temperature (red), exhibiting a zero-bias conductance dip. The minimum conductance $g_\mathrm{min}$, measured while waiting at $V_\mathrm{SD}=0$ for an extended time, is indicated as dashed line -- well below $g(V_{\rm SD}=0)$ when sweeping.}\vspace{-4mm}
\label{fig4}
\end{figure}

The CBTs consist of 7 parallel arrays with 64 angle evaporated tunnel junctions between metal islands\cite{Meschke04,Pekola94}, operating in the regime where the charging energy $E_\mathrm{C}$ is comparable to temperature\cite{Feshchenko13} (in contrast to the dot measurements in Fig.\,\ref{fig2} and \ref{fig3}, where $E_{\rm C}\gg k_B T_{\rm e}$). The full width at half minimum of the $V_{\rm SD}$ dependent conductance dip (see inset of Fig.\ref{fig4}) can serve as a primary thermometer\cite{Farhangfar97,Meschke11}. Here, however, we use the conductance dip $\Delta g/g$ as a secondary thermometer to avoid heating at finite bias\cite{Casparis2012}. At high temperatures where the CBTs are in thermal equilibrium with the refrigerator, $E_\mathrm{C}$ can be extracted from the temperature dependent $\Delta g/g$ using the relation\cite{Feshchenko13} $\Delta g/g=u/6-u^{2}/60+u^{3}/630$, where $u=E_\mathrm{C}/k_\mathrm{B}T_{\rm CBT}$. With $E_\mathrm{C}$ obtained in this way, the measured $\Delta g/g$ can be converted to the corresponding electron temperature $T_{\rm CBT}$ in the full temperature range.

Fig.\,\ref{fig4} shows $T_{\rm CBT}$ as a function of refrigerator temperature $T_{\rm MC}$ for two CBTs, measured using standard lock-in technique with AC excitation of $4\,\mu$V (experimentally chosen small enough to avoid heating effects). At the lowest temperatures, the zero-bias conductance $g(V_\mathrm{SD}=0)$ drops over a time span of several minutes and eventually saturates at a value $g_\mathrm{min}$, clearly lower than the minimum conductance from $V_\mathrm{SD}$ sweeps, see inset of Fig.\ref{fig4}. We therefore use $g_\mathrm{min}$ in order to obtain the conductance dip $\Delta g/g$ and consequently the electron temperature $T_{\rm CBT}$. The extracted electron temperatures $T_{\rm CBT}$ from both devices (circles and squares) agree very well with each other and with the refrigerator temperature $T_{\rm MC}$ down to about 15\,mK, where the electron temperatures begin to saturate, eventually reaching a minimum\cite{Casparis2012} $T_{\rm CBT}=$\,7.5\,$\pm$\,0.2\,mK for $T_{\rm MC}\approx$\,5\,mK.

We note that fitting the data in Fig.\,\ref{fig4} with a power law\cite{Casparis2012} $T_{\rm CBT}^p=T_S^p+T_\mathrm{MC}^p$ gives $p=2.7\pm0.2$ and $T_S=$\,6.9\,$\pm$\,0.1\,mK, where $T_S$ is a theoretical $T_{\rm MC}=$\,0\,K saturation limit of $T_{\rm CBT}$. This exponent $p$ is clearly lower than $p=5$ which corresponds to electron-phonon coupling\cite{pobell_1992}, thus indicating an additional cooling mechanism stronger than electron-phonon cooling at low temperatures, presumably electronic (i.e. Wiedemann-Franz\cite{pobell_1992}) cooling\cite{Casparis2012}. Nevertheless, if only electron-phonon coupling is assumed, the lowest electron temperature reached corresponds to a residual heat leak of 13\,aW per junction\cite{Meschke04,Casparis2012}.

In conclusion, we have presented miniature microwave filters uniting superior attenuation with efficient wire thermalization. The attenuation reaches $\geq$\,100\,dB above 150\,MHz or, when capacitors are added, already above 30\,MHz. The filters are modular, of small size, robust against thermal cycling, and the attenuation is well understood by skin-effect filtering in a lossy transmission line model. Combined with thermo-coaxes, one stage of Ag-epoxy filters reduces the electron temperature, measured with a GaAs quantum dot, from $\mathrm{\approx75\,mK}$ down to $\mathrm{\approx18\,mK}$. With an improved setup comprising a second Ag-epoxy filtering stage and a better chip holder, electron temperatures as low as 7.5\,$\pm$\,0.2\,mK are obtained in metallic CBTs.

Further experiments are required to determine whether the second filter stage or the improved chip carrier (or both) are necessary for the improvement from 18\,mK to 7.5\,mK. Finally, we point out that the filtering strategy introduced here was recently used to find evidence for helical nuclear spin order in GaAs quantum wires\cite{Scheller14} -- thus showing that the extensive filtering and thermalization introduced here can be important for studying new physics.

\begin{acknowledgments}
We thank J.\,P. Pekola for his support and G. Frossati, M. Steinacher, and A. de Waard for valuable inputs on the filter design. This work was supported by the Swiss Nanoscience Institute (SNI), NCCR QSIT, Swiss NSF, ERC starting grant, and EU-FP7 SOLID and MICROKELVIN.
\end{acknowledgments}


\providecommand{\noopsort}[1]{}\providecommand{\singleletter}[1]{#1}%

\end{document}